\documentclass[12pt]{article}

\usepackage{graphicx,bm}

\usepackage{multirow}

\usepackage{changes}

\usepackage{color}

\newcommand{\tced}[1]{\textcolor{black}{#1}}
\newcommand{\tcedd}[1]{\textcolor{black}{#1}}
\newcommand{\email}[1]{\footnote{#1}}
\newcommand{\affiliation}[1]{\\{}{\small{\em #1}} }

\begin{document}

 
\title{
Daejeon16 interaction with contact-term corrections for heavy nuclear systems
} 

\author{{Panagiota Papakonstantinou$^a$\email{ppapakon@ibs.re.kr (corresponding author)}, James P. Vary$^b$\email{jvary@iastate.edu}, Youngman Kim$^a$\email{ykim@ibs.re.kr}}
\affiliation{{$^a$Rare Isotope Science Project, Institute for Basic Science, Daejeon 34000, Korea}}
\affiliation{$^b$Department of Physics and Astronomy, Iowa State University, Ames, Iowa 50011, USA}
}

\date{\today}

\maketitle 

\begin{abstract}
The Daejeon16 two-nucleon interaction is employed in many-body approaches based on the mean-field approximation. 
The perturbative character of Daejeon16 is verified by comparing results  for $^{16}$O from the Hartree-Fock (HF) approximation and from the no-core shell model 
and by examining the magnitude of perturbative corrections to the HF energy in light and heavy nuclei. 
In order to approximately describe energies and radii across the nuclear chart, 
a phenomenological correction in the form of a two-plus-three-nucleon contact interaction is introduced. 
With fitted parameters we achieve a very good description of medium-mass nuclei in terms of energy and size and also in terms of the centroid energy of the giant monopole resonance and the dipole polarizability calculated within the random-phase approximation (RPA). 
Our results provide further justification for the use of Daejeon16 augmented with phenomenological corrections as an effective interaction of perturbative character in a variety of applications.  
\end{abstract}


\bibliographystyle{unsrt}

\section{Introduction\label{Sec:Intro}}

The non-relativistic nuclear quantum many-body problem traditionally consists in describing atomic nuclei and nuclear matter as aggregates of interacting nucleons in the framework of quantum mechanics~\cite{RS80,FW2003}. The degrees of freedom considered are not the fundamental ones of quantum chromodynamics (QCD), namely quarks and gluons, but inert nucleons in the non-perturbative regime of QCD. The relevant theoretical entities are the non-relativistic Hamiltonian (kinetic energy and potential operators) and the wave function. Solution of the Schroedinger equation for a realistic Hamiltonian should lead to an accurate description of known nuclear phenomena at low energies and reliable predictions for the properties of exotic nuclei. The above general concept, based on an intranucleon Hamiltonian with links to the underlying theory of QCD, roughly defines what we call the {\em ab initio} approach to nuclear structure. 

The complications are well-known. The Hamiltonian describing the strong nuclear force has a complex structure and includes strong repulsion at short distances as evident, for example, in two-nucleon scattering and indirectly in nucleon knock-out data~\cite{Arrington2012}. This suggests that the wave function will have high-momentum components or, equivalently, short-range correlations. The computational problem becomes challenging. Therefore, the many-body Schroedinger equation for non-homogeneous, strongly interacting, self-bound nuclei, except for the lightest nuclei, is computationally hard and requires controlled approximations to the exact many-body problem. 

One solution is the use of low-momentum nuclear Hamiltonians, where high momenta have been “integrated out”, formally by transforming a realistic nuclear interaction or by construction. 
Low-momentum interactions lead to improved convergence of results when employed in quantum many-body methods. 
These interactions are appealing for calculating bulk properties of the nuclear ground state (energy and size) and the nuclear response function to long-range operators using approximate many-body methods. 
Input realistic interactions include high-precision two-nucleon (NN) potentials such as the Argonne and Bonn families and chiral potentials~\cite{Machleidt2011}. 
These interactions must be accompanied by consistently defined three-nucleon (NNN) interactions. 
For convenience, we may refer to them as ``bare" interactions. 
Renormalization methods for generating low-momentum or ``effective" interactions from these bare interactions include 
the traditional $G-$matrix~\cite{Brueckner1955,Day1967}, 
$V_{\mathrm{low}-k}$~\cite{Bogner2003}, the unitary correlation operator method (UCOM)~\cite{Roth2010}, the similarity renormalization group (SRG)~\cite{Bogner2007}, and the Okubo-Lee-Suzuki method~\cite{Suzuki1994}. 
Thanks to such renormalization developments the {\em ab initio} program in nuclear physics has been thriving for several years with a wealth of successes in the description of light and medium-mass nuclei. 

Besides the genuine NNN and generally many-nucleon terms in the bare interactions, there are also induced many-nucleon interactions, which are generated by the renormalization methods. The presence of  many-nucleon forces considerably complicates the nuclear many-body problem and more so in the case of medium-mass and heavy nuclei. Therefore, there have been efforts to construct low-momentum or effective NN interactions such that the contribution of the many-nucleon terms is significantly reduced. For example, in the case of the UCOM the range of the tensor correlator was used early on as a free parameter adjusted such that no-core-shell-model (NCSM)~\cite{Navratil2000a,Navratil2000b} calculations employing only the two-nucleon part of the UCOM potential could reproduce the energies of light nuclei~\cite{Roth2005}. It turned out that binding energies throughout the nuclear chart could then be reproduced within many-body perturbation theory (MBPT), but nuclear radii were strongly underestimated~\cite{Roth2006}. 
 In addition, the energies of major giant resonances of closed-shell nuclei were overestimated in calculations within the random-phase approximation (RPA)~\cite{Paar2006}.
Further attempts employing SRG, with the renormalization flow parameter serving as a free parameter, typically led to overbinding in heavy nuclei. 
A combination of UCOM and SRG and the inclusion of a phenomenological repulsive correction in the form of a three-nucleon contact term was added 
to cure the above deficiencies to some extent~\cite{Gunther2010,Hergert2011,GPR2014}. 
In the same spirit, a phenomenological three-nucleon contact term of variable strength, depending on the nucleus and application, was added to the $V_{\mathrm{low}-k}$ or NNLO$_{\mathrm{opt}}$ potentials 
in calculations of the multipole response function within Tamm-Dancoff, RPA, and multiphonon approaches~\cite{Bianco2014,Knapp2014,Knapp2015}. 

On the one hand, such results suggest that a good description of light nuclei does not guarantee realistic saturation properties for nuclear matter. The issue has arisen also for chiral interactions, hence the subsequent development of chiral potentials with coupling constants adjusted also to heavier nuclei~\cite{Ekstrom2015}.
On the other hand, the NNN correction required to obtain realistic results in the above-mentioned examples was found roughly one order of magnitude weaker
than the analogous density-dependent terms needed to ensure saturation in the case of 
purely phenomenological effective interactions (such as Gogny and Skyrme)~\cite{Pap2020}. This suggests that much of the relevant physics is already present in the NN renormalized realistic interactions and the additional terms represent modest though non-negligible corrections.  

In this work we focus on the Daejeon16 interaction~\cite{Shirokov2016}, a relatively new and promising NN interaction with several applications already, especially in light nuclei. It was constructed from the Idaho N3LO interaction by applying an SRG evolution and finally a set of phase-equivalent transformations. The fitting process involved nuclei up to $^{16}$O.  Daejeon16 has found applications in the description of nucleon-nucleus scattering ~\cite{Shirokov2018a,Mazur2019a,Mazur2019b,Mazur2019c,Mazur2020}, the spectroscopy of light nuclei in favorable comparisons with the LENPIC NN+NNN interaction~\cite{Maris2019}, $sd-$shell nuclei as a valence interaction~\cite{Smirnova2019}, the tetraneutron resonance~\cite{Mazur2018,Shirokov2018b}, clustering~\cite{Rodkin2019} and in the development of artificial neural networks ~\cite{Negoita2019} -- see also \cite{Kim2019} for an overview of applications. 

The Daejeon16 interaction was developed to serve as a stand-alone NN interaction. In that sense it is a successor to the phenomenological JISP16~\cite{Shirokov2007} but provides superior convergence properties and performance in the case of  light nuclei~\cite{Shin2018,Kim2019}. Presently, we are interested in its perturbative behavior and ask whether it can be used as an effective interaction in the description of heavier nuclei within tractable many-body methods based on the mean-field approximation as a starting point. To this end, we first compare results for $^{16}$O from the Hartree-Fock (HF) approximation and from the NCSM and examine the magnitude of perturbative corrections. Comparisons with other interactions are made. In order to describe energies and radii of heavier nuclei we introduce a phenomenological correction in the form of a two-plus-three-body contact interaction. The implications for the saturation properties of Daejeon16 are discussed. Results for collective excitations are also discussed within the RPA. 

This paper is organized as follows. 
In Sec.~\ref{Sec:Elements} we provide for completeness the basic elements of the many-body methods and implementations used here, namely HF, MBPT and RPA.     
In Sec.~\ref{Sec:Data} we introduce the basic observables examined here and the corresponding experimental data. 
In Sec.~\ref{Sec:Pert} we verify the good convergence properties of Daejeon16. 
In Sec.~\ref{Sec:CSN} we present HF and MBPT results for heavier closed-shell nuclei and diagnose the need to introduce a phenomenological correction. 
Such a correction is determined and applied in HF and RPA calculations in Sec.~\ref{Sec:Pheno}. 
We conclude in Sec.~\ref{Sec:Concl}.

\section{Elements of Hartree-Fock, perturbation theory, and random-phase approximation\label{Sec:Elements}} 

The many-body methods used in this work, namely the Hartree-Fock (HF) approximation for the nuclear ground state, many-body perturbation theory (MBPT) 
for the correlation energy, and the random-phase approximation (RPA) for excited states have been standard tools of nuclear theory for a long time~\cite{RS80,Hey1994}. 
Here we review basic elements of the implementations employed for the purposes of the present study. 

HF serves as a basic variational method yielding the independent-particle many-fermion wavefunction (Slater determinant) which minimizes the total energy, i.e., the 
expectation value of the Hamiltonian. 
Here we consider the intrinsic $A-$nucleon Hamiltonian given in terms of the two-body intrinsic kinetic energy and a two-nucleon potential~\cite{Jaqua1992},  
\begin{equation} 
\hat{H} =   
 \sum_{i<j} \left( \frac{2}{Am}(\hat{\vec{p}}_i - \hat{\vec{p}}_j)^2  + \hat{V}_{ij} \right) ,
 \label{Eq:Hint}
\end{equation}  
in a straightforward notation. 
We comment that the NN interaction includes the Coulomb potential between protons. 
The HF equations are solved within a spherical harmonic-oscillator (HO) single-particle basis.  
Through an iterative procedure the ground-state wavefunction is obtained as a Slater determinant of occupied single-particle states (hole states), 
from which all basic properties of the ground state can be calculated.  
Energies and wavefunctions of the unoccupied states (particle states) are also obtained. 
Those can be used to compute perturbative corrections to the ground-state energy, which, to second order, are given by 
\begin{equation} 
\Delta E_{\mathrm{MBPT(2)}} = -\frac{1}{4}\sum_{p_1 p_2 h_1 h_2} \frac{|  \langle p_1p_2 | H | h_1 h_2 \rangle |^2}{ e_{p_1} + e_{p_2} - e_{h_1}  -  e_{h_2}   }, 
\end{equation} 
where the sum runs through all particle ($p_i$) and hole ($h_i$) states and $e_{a}$ is the HF single-particle energy of state $|a\rangle$.  

The HF solution serves also as a reference state for calculating properties of excited states within RPA. 
An external field formally represented by a single-particle operator $\hat{O}=\sum_{ij}O_{ij}a_i^{\dagger}a_j + h.c.$ is assumed to act on the nuclear ground state and create a phonon state 
\begin{equation} 
 | \nu \rangle = \sum_{ph} \left[ X_{ph}^{\nu} a_p^{\dagger}a_h - Y_{ph}^{\nu} a_h^{\dagger} a_p \right]  |g.s.\rangle  \, . 
\end{equation} 
Application of the quasi-boson approximation produces the standard RPA equations~\cite{Rowe1968}. 
Presently we use the self-consistent variant of RPA, i.e., the same Hamiltonian used to generate the HF solution, Eq.~(\ref{Eq:Hint}), 
is used also as a particle-hole interaction in RPA. 
All hole and particle states available from the HF solution, as determined by the original HO basis, are used to construct the $ph$ configuration space without any further energy truncation.  
The nuclear ground state is considered spherical and the excited phonons have good angular momentum and parity $J^{\pi}$. Details are provided in Ref.~\cite{Paar2006}.

\section{Observables and experimental data\label{Sec:Data}} 

We focus on bulk and mostly static properties of closed-shell nuclei.
In particular, we examine the ground-state energy per particle $E/A$, the point-proton radius $R_p$, the centroid energy of the isoscalar giant monopole resonance $E_c$(GMR), and the electric dipole polarizability $a_D$. 
The former three characterize the saturation point (energy, density, and compression modulus) of symmetric nuclear matter and the fourth characterizes the density dependence of the nuclear symmetry energy. In other words, these quantities can be used as proxies for bulk properties of nuclear matter.   
The $E_c$ and $a_D$ are defined in terms of energy moments of corresponding transition-strength distributions (excitation spectra) in the isoscalar-monopole and electric-dipole channel respectively.
The transitions are generated formally by acting on the nuclear ground state with the single-particle isoscalar monopole (ISM) and electric dipole (E1) operators 
\begin{equation} 
\hat{O}_{\mathrm{ISM}}=\sum_{i=1}^{A}r_i^2Y_{0}(\hat{r}_i) 
\end{equation} 
\begin{equation} 
\hat{O}_{E1}=\frac{N}{A}\sum_{p=1}^{Z}r_pY_{1}(\hat{r}_p) - \frac{Z}{A}\sum_{n=1}^{N}r_nY_{1}(\hat{r}_n) 
\end{equation} 
and the corresponding transition strength distributions are obtained by taking the transition matrix elements to excited states, 
\begin{equation} 
S(E) = \sum_f  |\langle f | \hat{O} | \mathrm{g.s.}\rangle |^2 \delta (E-E_f) .
\end{equation} 
The $k-$th moment of the transition strength distribution $S(E)$ is defined as
\begin{equation}
m_k = \sum_i E_i^kS(E_i) \,  ,     
\end{equation} 
where the sum runs over all excited states. 
The centroid energy of the distribution is defined via the first and zeroth moments, 
\begin{equation} 
E_c=m_1/m_0 .
\end{equation}
In the isoscalar monopole case almost all strength is exhausted by the giant monopole resonance. 
For more information see, e.g., Ref.~\cite{Garg2018}. 
\tced{Therefore, we calculate the centroid energy $E_c$(GMR) using the moments of the entire distribution. } 
The electric dipole polarizability is determined by the inverse energy weighted sum $m_{-1}$ of the electric-dipole strength distribution, in particular,  
\begin{equation} 
a_D = \frac{8\pi}{9} \frac{\hbar c}{137} m_{-1}(E1). 
\end{equation} 
For more information see, e.g., Ref.~\cite{RoP2018}. 

In practice, the ground-state energies and radii are calculated in the HF approximation and corrections are calculated with second-order perturbation theory~\cite{Roth2006}. 
Transition strength distributions are calculated within RPA~\cite{Paar2006,GPR2014}. 
For basic information on the current inplementations, see Sec.~\ref{Sec:Elements}.  

The data we use for comparison are collected in Table~\ref{Tab:Data}. 
Values for $E/A$ and for the charge root-mean-square radii $R_{\mathrm{ch}}$ are taken from the AME2016 evaluation~\cite{AME2016}. 
Point-proton radii $R_p$ are extracted from experimental measurements of $R_{\mathrm{ch}}$ for each nucleus $(A,Z)$ 
by applying, as in other studies~\cite{Ekstrom2015}, corrections which include  the Darwin-Foldy correction and effects of the finite nucleon size, 
\begin{equation} 
R_p^2 = R_{\mathrm{ch}}^2 - r_p^2 -\frac{N}{Z}r_n^2 - \frac{3\hbar^2}{4m_p^2c^2}, 
\label{Eq:Rp}
\end{equation}  
where $\frac{3\hbar^2}{4m_p^2c^2}=0.033$~fm$^2$, $r_p=0.8775(51)$~fm, and $r_n^2=-0.1149(27)$~fm$^2$. 
For convenience, we refer to these $R_p$ values as experiment (EXP) in what follows. 
Values  shown are rounded to the 4th significant digit at most, regardless of measurement precision, 
because high precision is superfluous for this work, as will become clear when examining the results.
Values for the $E_c$(GMR) are available from alpha-scattering experiments~\cite{Garg2018,Howard2020}, while measurements of complete dipole spectra in polarized proton scattering have made possible the extraction of $a_D$ values~\cite{Birkhan2017,Tamii2011,RoP2018}.

\begin{table} 
\begin{tabular}{l|cccccc} 
\mbox{~}
$\begin{array}{rl}  & \mbox{~}\!\!\!\!\mathrm{Nuclide} \\ \mathrm{Property} & \end{array} $        &  $^{16}$O  &     $^{40}$Ca  &  $^{48}$Ca  &  $^{90}$Zr  &     $^{132}$Sn & $^{208}$Pb \\
\hline
$-E/A$ [MeV]         & 7.976 &  8.551 & 8.667 &  8.710 &  8.355 & 7.867 \\
$R_{\mathrm{ch}}$ [fm]     & 2.699  &    3.478 & 3.477 &    4.264 &  4.709 & 5.501  \\
$R_p$ [fm]  & 2.581  &  3.387 & 3.393 &  4.199 & 4.650  & 5.450  \\
$E_c$(GMR) [MeV]$^{(\ast )}$  &  $--$ &  $20.2$  & $19.5$  & 18.13 &  $--$  & 13.96 \\
$a_D$  [fm$^3/e^2$]            &  $--$ &  $--$  &  $2.07(22)$  & $--$ &  $--$ & 20.1(6)  \\
\end{tabular}\vspace{-2mm}  
\flushleft{$^{(\ast )}$ Error bars for $E_c$(GMR) are of the order of $10^{-1}$MeV.}
\caption{Experimental data referenced in this work: 
ground-state energy per particle $E/A$~\cite{AME2016} (rounded), 
charge radius $R_{\mathrm{ch}}$~\cite{Angeli2013} (rounded), 
point-proton radius $R_p$ (see text), 
centroid energy of the giant monopole resonance $E_c$(GMR) (Ca isotopes: Ref.~\cite{Howard2020}; other: Ref.~\cite{Garg2018}), and electric dipole polarizability $a_D$~\cite{RoP2018}. 
In addition, we consider the energy per particle of $^{28}$O, 5.988~MeV, and of $^{100}$Sn, 8.253~MeV~\cite{AME2016}. 
A double dash ``$--$" indicates unavailable data or not used here.  \label{Tab:Data}}
\end{table} 

\section{Convergence properties of Daejeon16: $^{16}$O \label{Sec:Pert}} 

We first verify that the Daejeon16 interaction shows good convergence behavior with respect to the model space and in comparison with other interactions.
To this end, we compare various calculations available for $^{16}$O based on a selection of many-body methods and interactions. 
Specifically, for the purposes of the present study of the nucleus $^{16}$O, the following calculations were performed: 
\begin{itemize} 
\item 
HF calculations using the Daejeon16 interaction within a model space of 13 harmonic-oscillator shells ($e_{\max}=(2n+\ell )_{\max}=12$) with frequency $\omega_1 =20$~MeV$/\hbar$ (length parameter $b_1=1.44$~fm) and, for comparison, $\omega_2 =10$~MeV$/\hbar$ (length parameter $b_2=2.04$~fm). 
\item 
Perturbation-theory corrections, through second order, to the above results. 
\item 
HF calculations using the UCOM interaction~\cite{Roth2006}  within a model space of 13 harmonic-oscillator shells with $b=1.6$~fm corresponding to $\omega \approx 16$~MeV$/\hbar$. 
\item 
Perturbation-theory corrections, through second order, to the above HF results. 
\end{itemize} 
Hartree-Fock calculations represent here the ``zero-order" or mean-field level of approximation. 

Table~\ref{T:D16EnRp} shows the ground-state energy, point-proton r.m.s. radius of $^{16}$O and intrinsic kinetic energy (if known) obtained within different models and compared with data. 
Results are shown from HF calculations with the denoted interaction, followed by results from HF including many-body perturbation theory with the same interaction, if available (``+ MBPT"). 
No-core shell model calculations are also shown if available (``NCSM").  
The notation N3LO(x) is shorthand for the SRG-softened N3LO interaction with flow parameter $\lambda=x$~fm$^{-1}$.  
The notation MBPT(n) is short hand for up through n-th order perturbation theory. 
HF and MBPT results with the JISP16 and N3LO interactions are taken from Ref.~\cite{HuX2016}. 
The quoted NSCM results with JISP16 and Daejeon16 are taken from Refs.\cite{MaV2013} and \cite{Shirokov2016}, respectively. 
For the former case, a slightly different value, -145(8) MeV is quoted in Ref.~\cite{Shi2014}, but this is consistent with the result in Table~\ref{T:D16EnRp} to within their quoted uncertainties. 
\begin{table} 
\begin{center} 
\begin{tabular}{llccc}  
\hline \hline 
          &   $E$ [MeV]  &    $R_p$ [fm]    & $T_{\mathrm{int}}$ [MeV]    \\ 
\hline 
 Exp.    &   -127.619                     &                   2.581     &     \\ 
\hline 
\underline{Daejeon16}$^{(a)}$        & & & \\ 
 HF; $b_1$ ($b_2$) 
          &  { -106.5  (-106.5) }                  &    2.24   (2.24)     &   310 (310)  \\ 
+MBPT(2)
          &            {    -126.6 (-130.4) }                     &   2.36 (2.43)   &    \\ 
Difference 
          &             {   -20.1 (-23.9)  }                    &                     &      \\ 
NCSM
          &     -131.4(7)                 &     $\approx$ 2.4               &    290 \\ 
\hline 
\underline{UCOM}$^{(b)}$   & & & \\
 HF &   -56.3                       &    2.27                         &   305.6  \\ 
+MBPT(2)
& -67.1                  &        2.53                  &  \\ 
\hline  
\underline{JISP16}$^{(c)}$ & & & \\ 
 HF
          &   - 71.638                     &    1.791                               &        \\ 
 + MBPT(2) 
          &   -130.511   &            &   \\  
 + MBPT(3) 
          &    -134.771                    &    1.843                              &      \\ 
NCSM  
          &     -146(7)                      &   [2.0-2.2]                          &    [$>$350]    \\ 
\hline 
\underline{N3LO(x)}$^{(d)}$ & & & \\ 
 HF+N3LO(1.5) 
          &    -169.968                    &    2.031  & \\ 
 + MBPT(3)   &    -180.893                    &    2.042 & \\ 
 HF+N3LO(2.0) 
          &    -133.169                    &    2.029  & \\ 
 + MBPT(3)   &    -164.597                    &    2.040 &   \\  
 HF+N3LO(2.5) 
          &    - 85.173                    &    2.131  & & \\  
 + MBPT(3)   &    -149.419                    &    2.125 &   \\  
 HF+N3LO(3.0) 
          &     -44.102                    &    2.272  & & \\  
 + MBPT(3)   &    -139.767                    &    2.230  &  \\  
\hline 
\end{tabular} 
\end{center} 
\mbox{~}\hspace{-1.4cm} 
\begin{tabular}{l} 
$^{(a)}$ HF and MBPT: this work; NCSM: Ref.~\cite{Shirokov2016} \\
$^{(b)}$ From Ref.~\cite{Roth2006} \\
$^{(c)}$ HF and MBPT: Ref.~\cite{HuX2016}; NCSM: Ref.~\cite{MaV2013}\\
$^{(d)}$ From Ref.~\cite{HuX2016} 
\end{tabular} 
\caption{\label{T:D16EnRp} Ground-state energy and point-proton r.m.s. radius of $^{16}$O within different models and compared with data. 
For the kinetic-energy estimate $T_{\mathrm{int}}$ see text.
} 
\end{table} 

The total energy varies from model to model by 300\%. Note, however, that the total energy is the difference of two large quantities, namely the intrinsic kinetic energy $T_{\mathrm{int}}$ and  the potential energy $|V|$, which, separately, show smaller percentage variations. 
The HF kinetic energy (intrinsic) $T_{\mathrm{int}}$ for UCOM and Daejeon16 is approximately 305 and 
310~MeV, respectively.
This would mean that for these two HF calculations the HF potential energy varies by roughly 15\%. 
The $T_{\mathrm{int}}$ values from the other calculations are not available, but from the conjugate relation between momentum and distance we expect that $T_{\mathrm{int}}$  is anticorrelated with the nuclear size represented here by the proton radius $R_p$. 
If we assume roughly that $T_{\mathrm{int}}\propto R_p^{-2}$ we find that 
the HF potential energy $V$ for the tabulated HF calculations 
varies between approximately -350 and -560 MeV only.

We conclude from the results in Table~\ref{T:D16EnRp} that Daejeon16 provides faster convergence than JISP16 since the magnitude of the second-order perturbative corrections relative to the HF energy 
is smaller by about a factor of 3. Furthermore, the HF+MBPT(2) result for Daejeon16 is closer to its NCSM result, with a difference lower than $\approx 5$~MeV, 
compared to the corresponding difference of 15~MeV for JISP16. 
We proceed to examine the perturbative corrections also in heavier nuclei.

\section{Results for closed-shell nuclei\label{Sec:CSN}} 
Having confirmed the perturbative behavior of Daejeon16 in $^{16}$O we proceed to test this behavior also in heavier nuclei and to examine the results in comparison with experimental data. 
 
\begin{figure}[htb]
 \centering
\includegraphics[width=0.8\columnwidth]{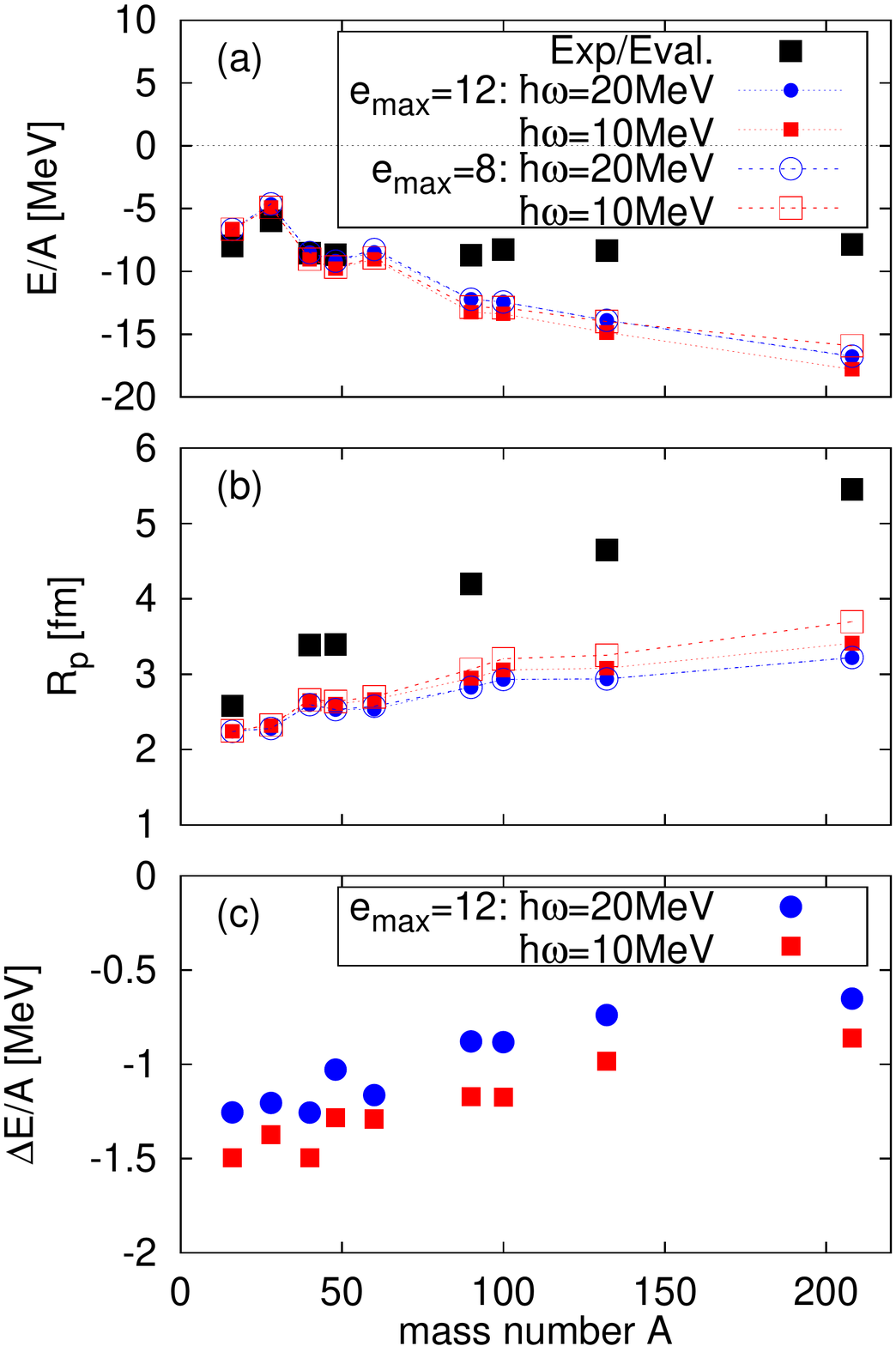}\\
\caption{Ground-state properties obtained with the Daejeon16 interaction for 
$^{16}$O, 
 $^{28}$O, 
$^{40}$Ca, 
$^{48}$Ca, 
$^{60}$Ca, 
$^{90}$Zr, 
$^{100}$Sn, 
$^{132}$Sn, 
$^{208}$Pb.
Results for various harmonic-oscillator bases are shown to examine convergence. 
Lines are drawn solely to guide the eye. 
(a) HF results for $E/A$. (b) HF results for $R_p$.  (c) Perturbative correction to the energy (note the expanded scale).
\label{Fig:HFd} 
}
\end{figure}

Results for the ground-state energies per nucleon and proton radii of various nuclei obtained within Hartree-Fock for the indicated harmonic oscillator bases (frequency and $e_{\max}$)  
compared with experimental or recommended values (Table~\ref{Tab:Data}) 
are shown in Fig.\ref{Fig:HFd}(a),(b). 
We note that $^{28}$O is correctly predicted particle-unbound at the HF level, i.e., the neutron Fermi energy is found positive, while for $^{60}$Ca the Fermi energy is correctly predicted negative when the more-optimal value $\hbar\omega =10~$MeV is used.  
Fig. \ref{Fig:HFd}(c) shows the correction to the energy per nucleon coming from second-order perturbation theory. 

$^{16}$O is found somewhat underbound within HF using Daejeon16 as already realized. 
Perturbative corrections, as we have seen, lower the energy of $^{16}$O to within about 0.1~MeV/$A$ of the experimental result. 
However, experimental data and HF calculations with Daejeon16 cross around Ca and heavier nuclei are found increasingly overbound. 
Since HF is a variational method within the simple model space of Slater determinants, an extended Hilbert space accounting for more correlations would lower the calculated energy further. 
Similarly, we find that these nuclei are more bound when the second-order perturbative corrections of Fig.~\ref{Fig:HFd}(c) are included. 

Interestingly, as seen in Fig.~\ref{Fig:HFd}(c), the perturbative energy correction per particle is about the same for all nuclei with a magnitude of roughly $0.5-1.5$~MeV. 
This value is quite small compared with the perturbative corrections found with other interactions (see, e.g., \cite{Roth2006}). 

As shown in Fig.~\ref{Fig:HFd}(b), the point-proton rms radii are found to be small compared with experiment for heavier nuclei. 
In the case of the charge radius of $^{16}$O the extended model space of the
NCSM can provide a correction of more than 0.1fm (Table~\ref{T:D16EnRp}) in the 
direction of agreement with experiement. Corrections to $R_p$ for heavier nuclei from perturbation theory would likely be similar in size, as
the results reported in Ref.~\cite{Roth2006} suggest and as we confirmed for Ca isotopes (not shown). 
In that case they would not be
sufficient to compensate for the discrepancies observed with our HF results compared with data. 

\begin{figure}[htb]
\mbox{$~$}\hspace{5mm}  
\includegraphics[width=0.8\columnwidth]{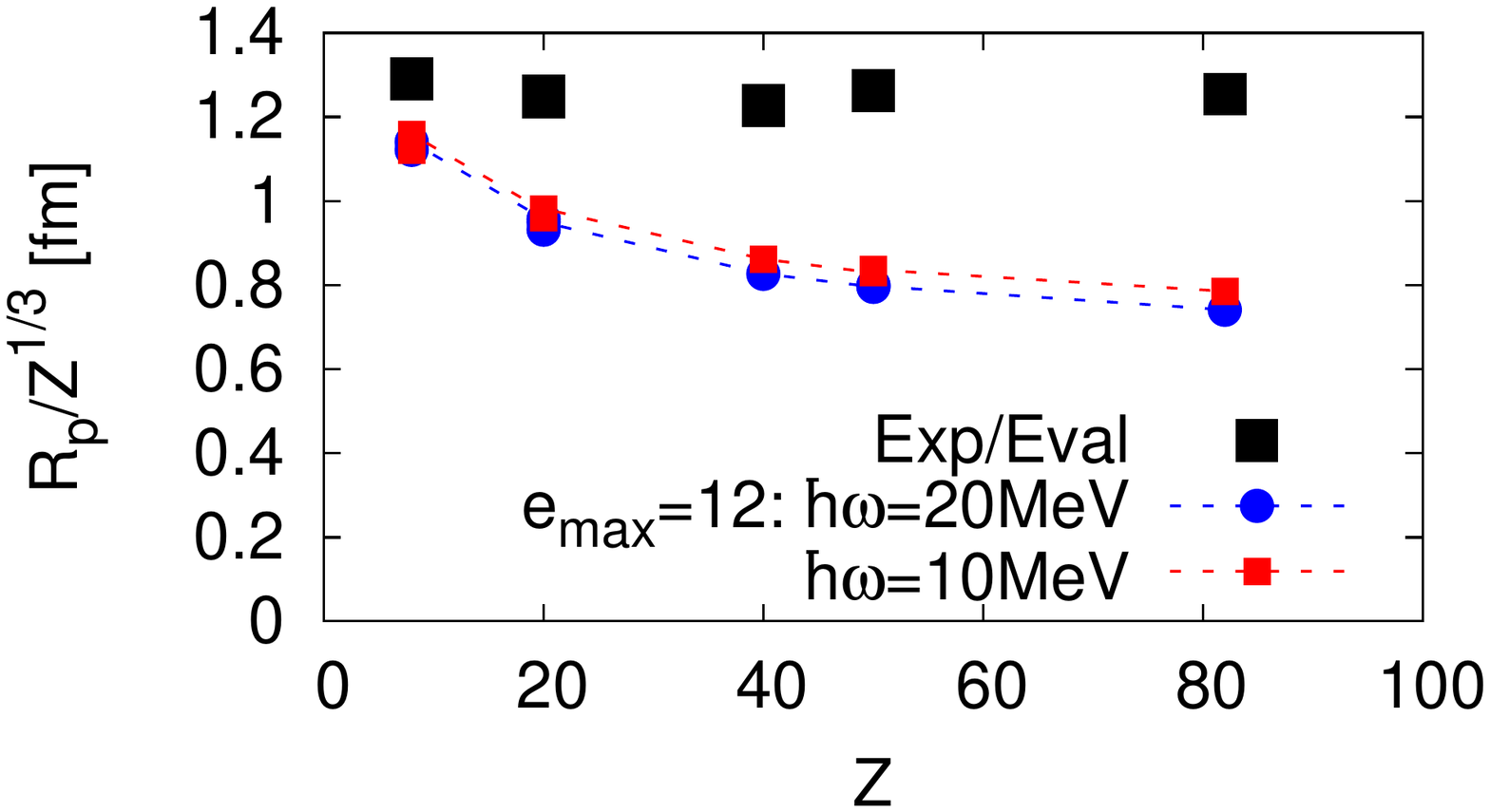}\\[-65mm]
\caption{Proton radius as a function of the atomic number $Z$. The empirical value divided by $Z^{1/3}$ is roughly stable 
demonstrating saturation. 
By contrast, the calculated radii divided by the cubic root of $Z$ continue to decrease with increasing $Z$.  
\label{Fig:r0} 
}
\end{figure}

We can infer from the results shown in Fig.~\ref{Fig:HFd} that the main issue is the saturation property of Daejeon16. 
The issue is less relevant for 
$^{16}$O which is light enough to be less sensitive to Daejeon16's saturation properties. 
However, the addition of nucleons (heavy nuclei) leads to overbound and denser systems compared with experiment. 
In the HF approximation, we attain a density higher than 0.6~fm$^{-3}$ inside $^{208}$Pb. 
The situation is further illustrated in Fig.~\ref{Fig:r0}. 
The empirical value for the charge radius divided by $Z^{1/3}$ stabilizes to a constant value demonstrating saturation. 
By contrast, the calculated radii divided by $Z^{1/3}$ continue to decrease in heavy nuclei. 
The trend in Fig.~\ref{Fig:r0} suggests possible stabilization in heavy systems. 

To summarize, Daejeon16 is a perturbative interaction and heavy nuclei are predicted to exist with finite though excessive density. 
Next we search for a phenomenological correction to Daejeon16 that will greatly reduce the differences of these results from experimental data.  
Observing in Fig.~\ref{Fig:HFd}(a) that the HF energy is minimized for $\hbar\omega=10$~MeV in the heavy systems, 
we will now work only with this value for $\hbar\omega$ and with $e_{\max}=12$.

\section{Phenomenological correction~\label{Sec:Pheno}}

In order to describe heavier nuclei with the Daejeon16 interaction additional repulsion must be introduced. 
It should be active especially in the bulk of nuclei.  
The desired effect could be introduced via a phenomenological correction to the Hamiltonian 
in the form of a contact three-nucleon repulsive term, 
\begin{equation} 
V_3 (i,j,k) = t_3\delta  (\vec{r}_i-\vec{r}_j) \delta (\vec{r}_j-\vec{r}_k).   
\label{Eq:Vijk} 
\end{equation} 
On the mean-field level the above interaction is equivalent to a two-nucleon interaction with a density-dependent coupling strength~\cite{Tsai1978}  
\begin{equation}
V_{2}(i,j) = \frac{t_3}{6}(1+\hat{P}_{\sigma})\rho ([\vec{r}_i+\vec{r}_j]/2)\delta (\vec{r}_i-\vec{r}_j) ,  
\label{Eq:dV22} 
\end{equation}
which is a special case of  
\begin{equation}
V_{2}(i,j;x_3,\alpha) = \frac{t_3}{6}(1+x_3\hat{P}_{\sigma})\rho^{\alpha}([\vec{r}_i+\vec{r}_j]/2)\delta (\vec{r}_i-\vec{r}_j) ,  
\label{Eq:dV2} 
\end{equation} 
with $x_3=1$ and $\alpha =1$. In the above, $\hat{P}_{\sigma}$ is the spin exchange operator for antisymmetrization and $\rho (\vec{r})$ is the local nucleon density.
We adopt here the familar notation of Skyrme functionals.  
If we set $\alpha=0$, expression (\ref{Eq:dV2}) gives the equivalent of a contact two-body interaction 
\begin{equation}
 \frac{t_3}{6}(1+x_3\hat{P}_{\sigma})\delta (\vec{r}_i-\vec{r}_j).  
\label{Eq:dVcont} 
\end{equation}

Within HF, we found that a single correction term of the above form for all values of $x_3$ and $\alpha$ fails to yield reasonable results simultaneously for the energy and radius of nuclei. 
The correction required for the calculated radius to be realistic is so strong that nuclei become unbound 
in the HF approximation using our chosen basis.
Noting that 1) the radius requires a correction in the bulk of the nucleus (near saturation density), 2) the energetics of nuclei are largely determined by the surface nucleons (subsaturation densities) and 3) the energetics of light nuclei such as  $^{16}$O are already rather optimal without a phenomenological correction, we deduce that an attractive counterterm dominating at low densities is needed such as to offset the excess repulsion introduced to the nuclear energies. 
Therefore we explore 
a phenomenological correction defined by a density-dependent two-nucleon potential of the form 
\begin{equation} 
V_{\rho}=\frac{1}{6}(1+\hat{P}_{\sigma})\{t_0 +  t_3\rho ([\vec{r}_i+\vec{r}_j]/2) \}\delta (\vec{r}_i-\vec{r}_j) 
\label{Eq:t0t3} 
\end{equation} 
with $t_0<0$ and $t_3>0$. We introduced the symbol $t_0$ corresponding to six times the familar $t_0$ term of a Skyrme functional. 
On the mean-field level this corresponds to a correction to the energy per particle of homogeneous nuclear matter equal to~\cite{Vautherin1972} 
\begin{equation} 
\Delta E_{\rho}/A = \frac{1}{16}(t_0 + t_3\rho )\rho (1-\delta^2) \, , 
\label{Eq:DESNM} 
\end{equation} 
where $\delta=(\rho_n-\rho_p)/\rho$ is the isospin asymmetry.  
An advantage of the linear form (\ref{Eq:t0t3}) of density dependence, as opposed to a fractional-power dependence, is that it corresponds to a true two-plus-three-nucleon contact interaction. 
Therefore, if successful, it could be introduced also in extended many-body approaches (contigent to defining an appropriate momentum cut-off).

\begin{figure}
  \centering
\includegraphics[width=0.7\columnwidth]{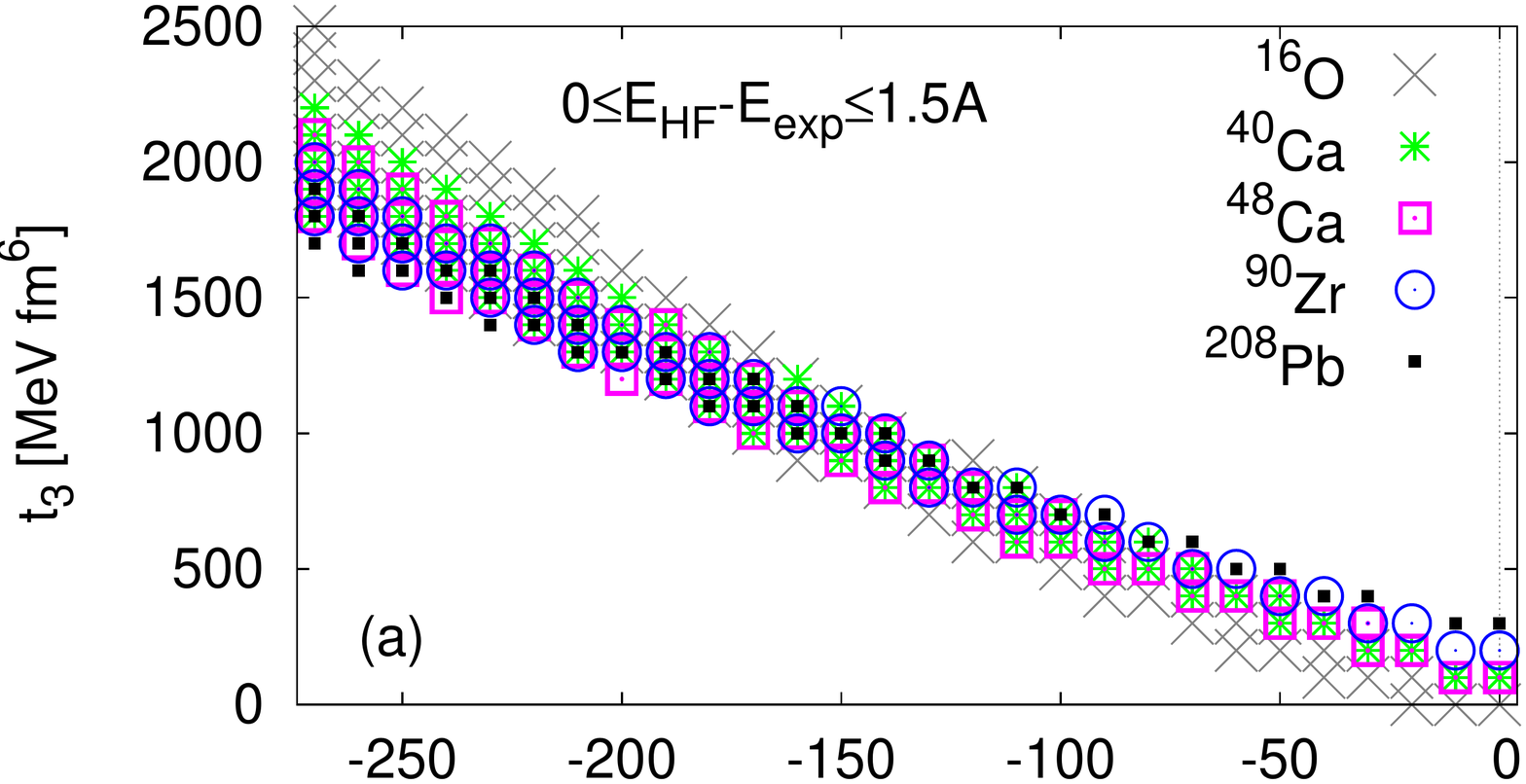}\\[-25mm] 
\includegraphics[width=0.7\columnwidth]{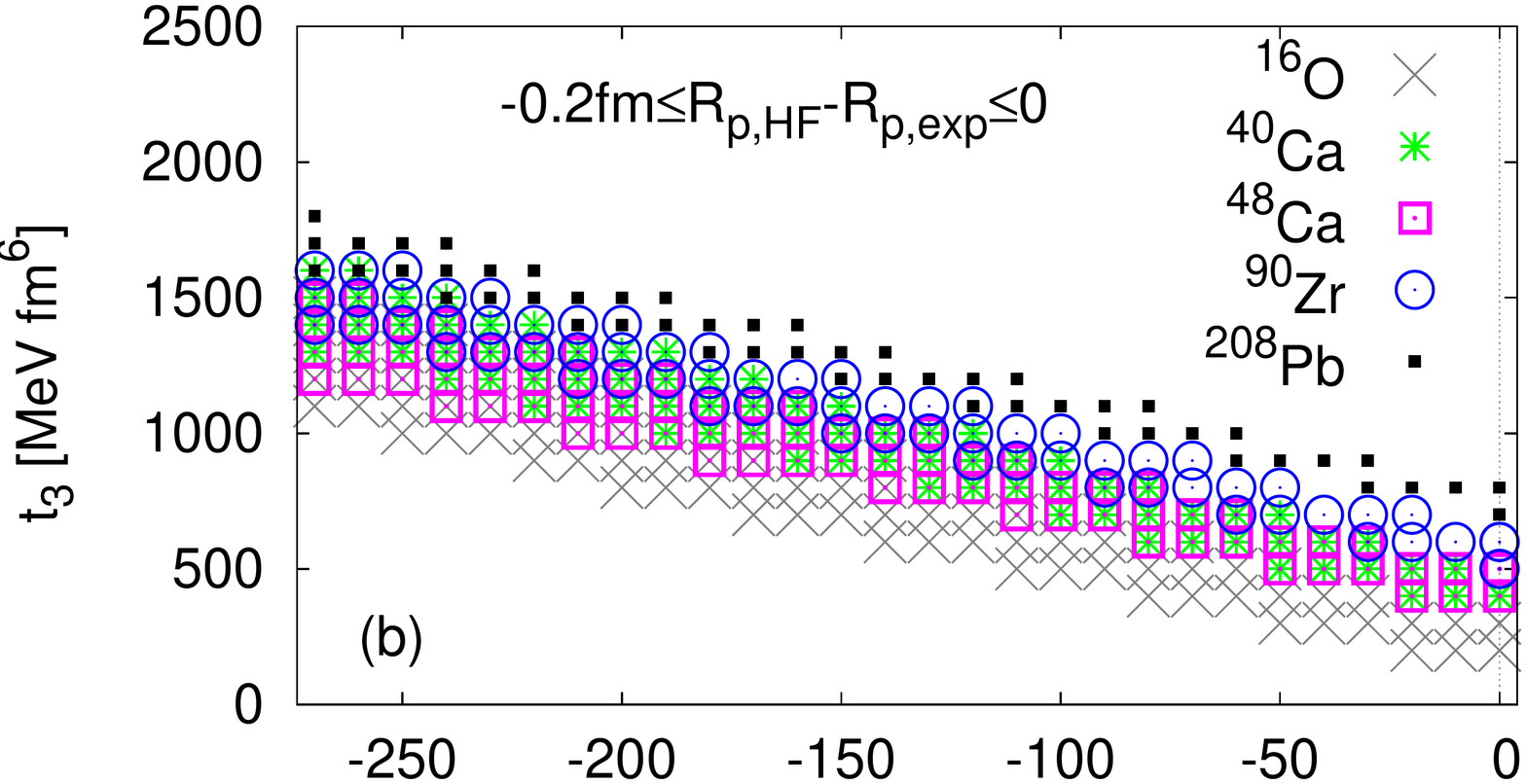}\\[-25mm] 
\includegraphics[width=0.7\columnwidth]{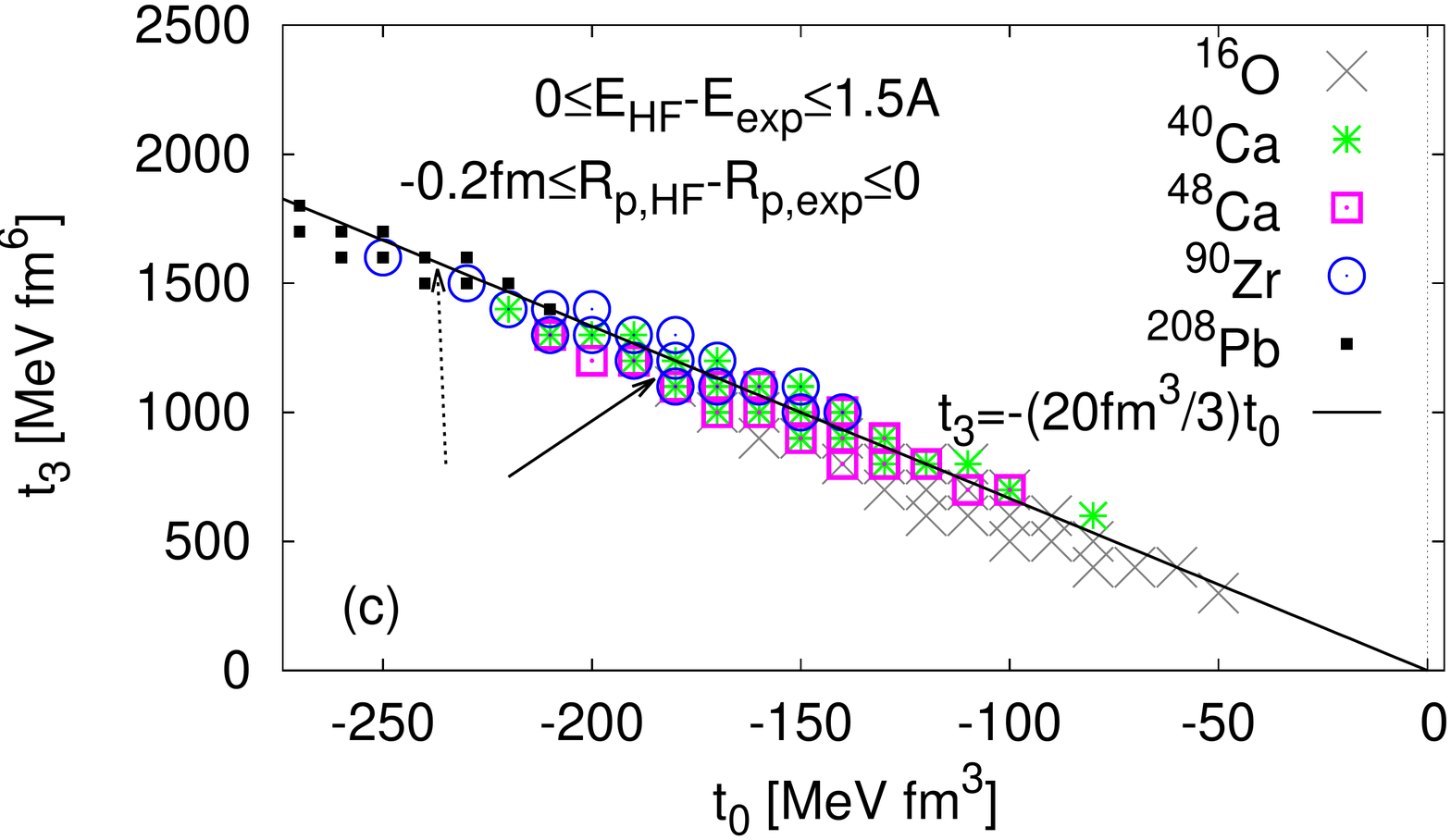}\\[-10mm] 
  \caption{Acceptable combinations of $(t_0,t_3)$ values for the correction term Eq.~(\ref{Eq:t0t3}) based on criteria (\ref{Eq:EToler}),~(\ref{Eq:RToler}) for HF results for the displayed nuclei.
(a) Applying only criterion (\ref{Eq:EToler}).  
(b) Applying only criterion (\ref{Eq:RToler}).   
(c) Applying both criteria. The arrows point to potentially optimal regions on the parameter plane.
}
  \label{Fig:ERToler}
\end{figure}

In principle one could fit the two parameters to the masses and radii of selected nuclei. 
For reasons that will be clarified below we prefer to choose the best values by inspecting the behavior of the results on the $t_0-t_3$ parameter plane. 
Let us consider as satisfactory outcomes those HF solutions for which the energy per particle is higher than or equal to the experimental value, the deviation being no more than 1.5MeV per particle 
\begin{equation} 
  0 \leq (E_{\mathrm{HF}} - E_{\mathrm{exp}}) \leq  1.5~A~\mathrm{MeV} \, .
\label{Eq:EToler} 
\end{equation} 
In a similar spirit, we adopt a condition for the proton radius 
\begin{equation} 
 - 0.2~\mathrm{fm} \leq R_{p,\mathrm{HF}} - R_{p,\mathrm{exp}} \leq 0  \,. 
\label{Eq:RToler}
\end{equation} 
The tolerance values of 1.5~MeV and 0.2~fm are based on the sizes of the corrections obtained from MBPT for Daejeon16 discussed above. 
We have performed HF calculations for values of $t_0$ from -270~MeV~fm$^3$ to 0 in steps of 10~MeV~fm$^3$  and for values of $t_3$ from 0 to {2500}~MeV~fm$^6$ 
in steps of 100~MeV~fm$^6$. 
In Fig.~\ref{Fig:ERToler}(a) and Fig.~\ref{Fig:ERToler}(b) we show  the acceptable sets of parameters for each of the examined nuclei based on the energy criterion (\ref{Eq:EToler}) 
and the radius criterion (\ref{Eq:RToler}) separately.   
In Fig.~\ref{Fig:ERToler}(c) we show the acceptable sets of parameters if we enforce both criteria simultaneously. 
%
From these results, we observe the following: 
\begin{itemize} 
\item
For each nucleus there is a band of $(t_0,t_3)$ values reproducing the desired energy. 
The bands are roughly linear but each with different slope. 
The region where they all overlap is substantial but lies away from zero, centered approximately at $t_0\approx -160$~MeV~fm$^3$ and $t_3\approx 1100$~MeV~fm$^6$. 
It is clear that a single parameter ($t_0$ or $t_3$) as in Eq.~(\ref{Eq:dV2}) would indeed fail to give satisfactory results.   
\item 
Radii cannot be simultaneously described for all nuclei: for given $t_0$ heavier nuclei require a stronger repulsion $t_3$  than do lighter nuclei. 
\item 
When the tolerance criteria are enforced for both the energy and the radius the region of potentially acceptable $(t_0,t_3)$ values shrinks considerably. 
Although it is still nucleus-dependent, it corresponds to a narrow and roughly linear band, 
described approximately by the relation  
\begin{equation} 
t_3 \approx (-20~\mathrm{fm}^3/3)t_0 \, .
\label{Eq:t0t3lin}
\end{equation} 
\end{itemize} 
Two small regions are identified by arrows on Fig.~\ref{Fig:ERToler}(c) as potentially optimal: 
The rightmost arrow points to a region where all but $^{208}$Pb could be satisfactorily descibed while the leftmost one to a region where heavier nuclei including $^{208}$Pb could be satisfactorily described, but not $^{16}$O and $^{40,48}$Ca.  

\begin{figure}
  \centering
 \includegraphics[width=0.8\columnwidth]{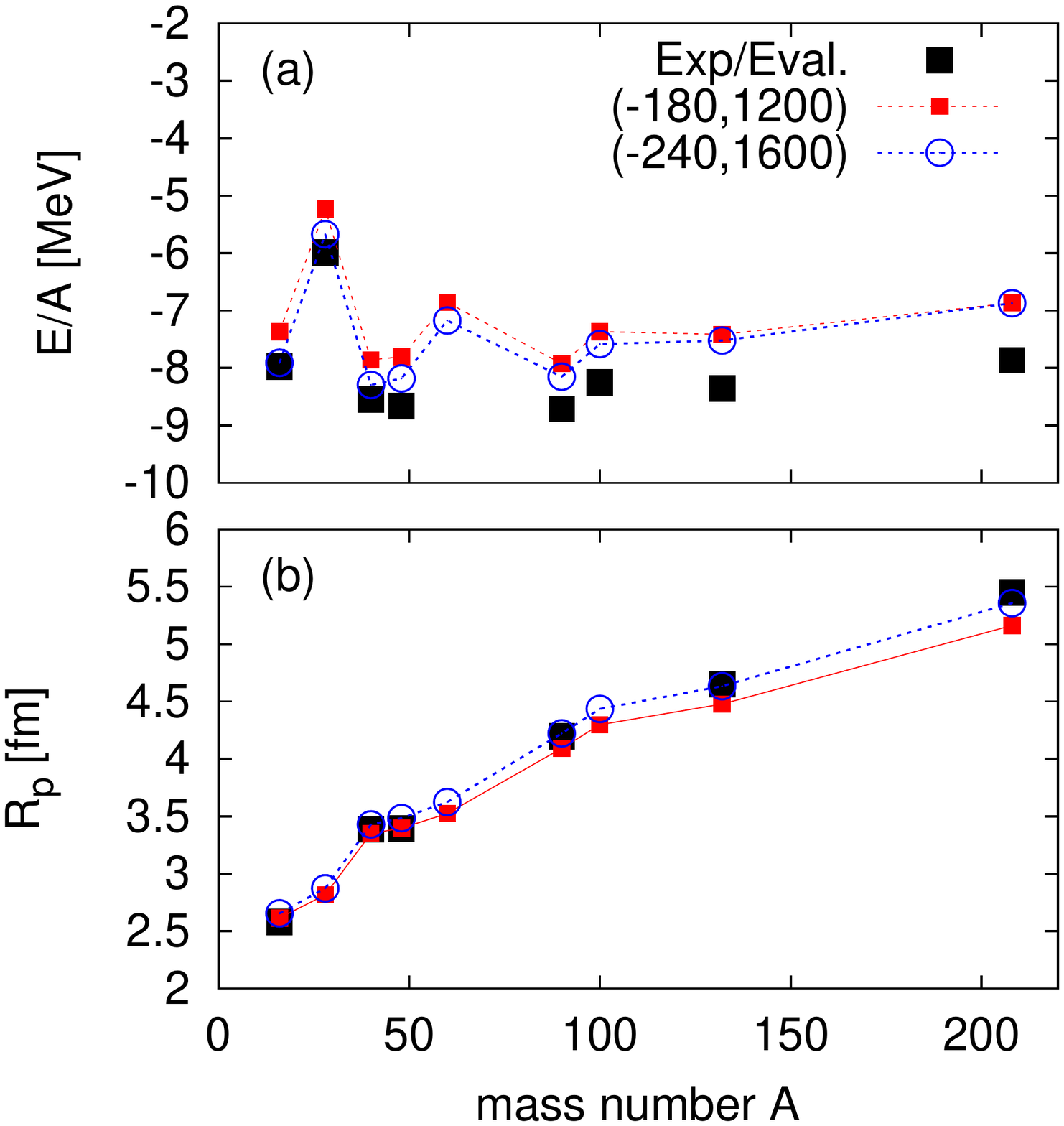}\\[-35mm]
  \caption{HF ground-state properties obtained with the Daejeon16 including correction (\ref{Eq:t0t3}) for the optimal $(t_0,t_3)$ values identified in Fig.~(\ref{Fig:ERToler})(c) by the arrows and indicated in units of $($MeV~fm$^3,$~MeV~fm$^6)$ 
for the nuclei 
$^{16}$O, 
 $^{28}$O, 
$^{40}$Ca, 
$^{48}$Ca, 
$^{60}$Ca, 
$^{90}$Zr, 
$^{100}$Sn, 
$^{132}$Sn, 
$^{208}$Pb.
(a) $E/A$. (b) $R_p$.
Experimental data are also shown where available. 
}
  \label{Fig:FinER}
\end{figure}

We conclude that no optimal pair of parameters exists for describing all examined nuclei from $^{16}$O to $^{208}$Pb 
within the dual criteria of Eqs.~(\ref{Eq:EToler}) and (\ref{Eq:RToler}).  A single global fit would require relaxing at least one of our criteria. 
However, depending on the application, for example if we are interested in a specific region of the nuclear chart, 
a pair of parameters along the band depicted on Fig.~\ref{Fig:ERToler}~(c) could be useful. 
As an illustration we show in Fig.~\ref{Fig:FinER} and tabulate in Table~\ref{Tab:ERpCorr} the energy and radius of closed-shell nuclei obtained with two sets of values $(t_0,t_3)$ in the regions indicated 
on Fig.~\ref{Fig:ERToler}(c) namely $(-180$~MeV~fm$^3,1200$~MeV~fm$^6)$ and$(-240$~MeV~fm$^3,1600$~MeV~fm$^6)$.
 They are confirmed as potentially appropriate corrections for mid-mass and heavy nuclei, respectively\tcedd{, assuming additional perturbative corrections can be included.}   
In addition, they provide a rather good description of light nuclei at the level of HF. 
We note that the neutron Fermi energy for $^{60}$Ca is predicted negative with the selected sets of parameters, while for $^{28}$O it changes sign at about $t_0=-167$~MeV~fm$^3$ along the line shown in Fig.~\ref{Fig:ERToler}(c) (Eq.~(\ref{Eq:t0t3lin}).) Finally,   
 it is worth noting that the corrections represented by the above $t_3$ values are weaker than those required in the case of SRG-evolved, 
$V_{\mathrm{low}-k}$ and NNLO$_{\mathrm{opt}}$ potentials with $t_3$ values from 2000 to 5000 MeV~fm$^6$~\cite{Gunther2010,Hergert2011,Bianco2014,Knapp2015}.   
%
%

\begin{table} 
\begin{tabular}{|c|cc|cc|}
\hline
             &  {}\,\,\,\, $E/A$ [MeV]                       &           &             \,\,\,\, $R_p$ [fm]  &  \\ 
             &  (-180,1200)  (-240,1600)  &    Exp.       &                (-180,1200)  (-240,1600)  &   Exp. \\
\hline
$^{16}$O    &  -7.367  \,\,\,\,\, \,\, -7.912 & -7.976      &          2.614 \,\,\,\,\,\, \,\,2.655  & 2.581 \\  
$^{28}$O    &  -5.235  \,\,\,\,\,\,\, -5.673 & -5.988     &          2.817 \,\,\,\,\,\,\,\, 2.874 &   $-$ \\ 
$^{40}$Ca  &  -7.861  \,\,\,\,\,\,\, -8.297 & -8.551    &          3.350 \,\,\,\,\,\, \,\,3.429 &  3.387  \\ 
$^{48}$Ca  &  -7.806  \,\,\,\,\,\,\, -8.184 & -8.667     &          3.396 \,\,\,\,\,\,\,\, 3.486 &  3.393 \\ 
$^{60}$Ca  &  -6.856  \,\,\,\,\, \,\,-7.174 & $-$     &          3.525 \,\,\,\,\,\, \,\,3.624 &   $-$    \\ 
$^{90}$Zr   &  -7.927  \,\,\,\,\,\,\, -8.155 &  -8.710    &          4.092 \,\,\,\,\,\,\,\, 4.223&  4.199  \\ 
$^{100}$Sn & -7.368  \,\,\,\,\,\,\, -7.586  &  -8.253   &          4.299 \,\,\,\,\,\,\,\, 4.436 & $-$  \\ 
$^{132}$Sn & -7.415  \,\,\,\,\,\,\, -7.522  & -8.355  &          4.478 \,\,\,\,\,\,\,\, 4.634 & 4.650  \\ 
$^{208}$Pb & -6.867  \,\,\,\,\,\,\, -6.871  &  -7.867   &          5.163 \,\,\,\,\,\, \,\,5.358 & 5.450  \\ 
\hline 
\end{tabular} 
\caption{The theoretical results plotted in Fig.~\ref{Fig:FinER} compared with data from experiments or evaluations. For references to data see Table~\ref{Tab:Data}. \label{Tab:ERpCorr}}  
\end{table} 

Applying Eq.~(\ref{Eq:DESNM}) to isospin-symmetric matter ($\delta=0$) and using the relation (\ref{Eq:t0t3lin}), we find that the correction to the energy per particle of  homogeneous symmetric matter can now be recast as 
\begin{equation} 
\Delta E_{\rho}/A \approx \frac{t_0}{16} \left(1 - \frac{\rho}{0.15~\mathrm{fm}^{-3}} \right) \rho  
\end{equation} 
and vanishes for density $\rho \approx 0.15$~fm$^{-3}$. 
Thus the effect of the correction is attractive at subsaturation densities and repulsive near and above saturation density, as anticipated. 
It is worth noting that, although the phenomenological correction is necessary, it is also quite small for subsaturation densities: for the sets of optimal values discussed above the energy correction amounts to a few hundereds keV per particle, to be compared with -16~MeV per particle, the empirical energy at saturation. 
On the other hand, it becomes large at suprasaturation densities. 
If we set $\rho=0.5$~fm$^{-3}$  we obtain a correction of more than $10$ MeV per particle. 
The interior density of heavy nuclei calculated within HF without the correction exceeds $0.5$~fm$^{-3}$, while it reaches more realistic values when the correction is included. 
As a result, the inclusion of the correction term leads to dramatic changes in the results for finite nuclei.

\begin{figure}
  \centering
 \includegraphics[width=0.8\columnwidth]{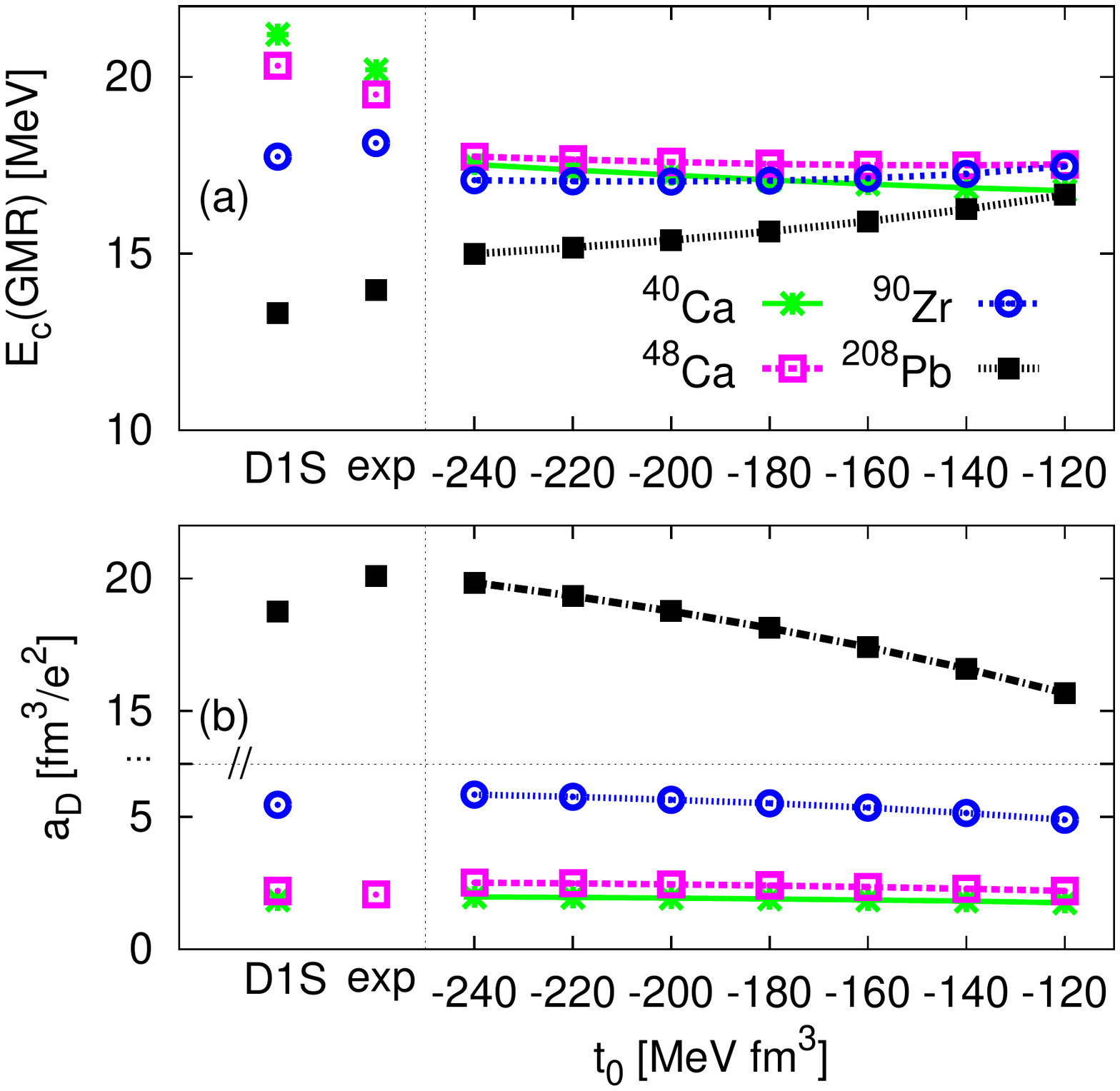}\\[-35mm]
  \caption{(a) GMR centroid energy and (b) electric dipole polarizability of $^{40}$Ca, $^{48}$Ca, $^{90}$Zr, $^{208}$Pb, calculated within RPA with Daejeon16 including correction (\ref{Eq:t0t3}) for the shown values of $t_0$ and for $t_3=(-20$~fm$^3/3)t_0$,  compared with RPA results with the Gogny D1S interaction (``D1S") and experimental data (``exp") where available (see Table~\ref{Tab:Data}). 
}
  \label{Fig:EcaD}
\end{figure}

Finally we examine $E_c$(GMR) and $a_D$ by calculating the monopole and electric-dipole spectra within the RPA formalism.    
The HF basis is used with no truncation other than that already imposed by the HO basis of $e_{\mathrm{max}}=12$. 
The necessary rearrangement terms for the density-dependent interaction are included as described in Ref.~\cite{Giambrone2003}. 
RPA calculations have been performed for
\begin{equation} 
t_0=-120, -140, \ldots , -240~ \mathrm{MeV~fm}^3
\end{equation} 
and $t_3=(-20~\mathrm{fm}^3/3)~t_0$. 
To put the results into perspective we compare them with RPA results obtained with the standard phenomenological interaction Gogny D1S. 
\tced{ 
Results with the Gogny D1S interaction were obtained with the same RPA model based on a HF reference state. 
As in previous works employing the D1S interaction~\cite{Hergert2011,GPR2014}, a $14\hbar\omega$ HO single-particle basis is used here and the HO length parameter for each nucleus 
has been chosen such as to minimize the intrinsic Hamiltonian's expectation value in the ground-state. 
}
Results are depicted in Fig.~\ref{Fig:EcaD} \tced{and tabulated in Table~\ref{Tab:EcaD}} for $^{40}$Ca, $^{48}$Ca, $^{90}$Zr, and $^{208}$Pb. 
The results for the lighter nuclei are very weakly affected by the correction terms, confirming the relative weakness of the latter at subsaturation densities. 
The results for the GMR centroid, Fig.~\ref{Fig:EcaD}(a), are best for the heavier nuclei $^{90}$Zr and $^{208}$Pb, 
suggesting a realistic description of the SNM compressibility near the saturation point but less so at lower densities.  
In the case of the dipole polarizability, Fig.~\ref{Fig:EcaD}(b), the performance of Daejeon16 supplemented with stronger correction terms ($t_0\approx-240$~MeV~fm$^3$) 
is very good for both $^{48}$Ca and $^{208}$Pb, suggesting realistic isovector properties. 
%
%
\begin{table}
\centering
\begin{tabular}{|c|cccc|cccc|}
\hline 
 &  $E_c${\small{(GMR)}}&\mbox{~}\hspace{-12mm}[MeV]                             &                    &              \\
                       &  $^{40}$Ca            &  $^{48}$Ca                        &  $^{90}$Zr &  $^{208}$Pb   \\
\hline    
  $ (-180,1200)$ & 17.08     & 17.54       &  17.06        &     15.62    \\ 
  $(-240,1600)$ & 17.52      & 17.75       &  17.08       &  14.99         \\ 
        Exp.     &  20.2(1)        &   19.5(1)   &   18.13(9)  &  13.96(20)  \\ 
        D1S      &   21.21          & 20.32      &  17.74         &  13.32        \\ 
\hline
\hline 
 &   \mbox{~}\,\,\,$a_D$  &\hspace{-12mm}[fm$^3/e^2$]       &                      & \\
                    &  $^{40}$Ca           &  $^{48}$Ca           &  $^{90}$Zr   &  $^{208}$Pb \\
\hline    
  $ (-180,1200)$ 
                          &     1.90    &   2.41       & 5.51     &   18.1        \\ 
  $(-240,1600)$   
                          &  1.98  &  2.52  &  5.84  &  19.8  \\      
        Exp.   
                        &  $-$  &  2.07(22)  & $-$  &  20.1(6)  \\    
        D1S    
                          &  1.86  & 2.20 &  5.45  &  18.7  \\
\hline
\end{tabular} 
\caption{
Theoretical values of $E_c$(GMR) and $a_D$ obtained within RPA using Daejeon16 and correction (\ref{Eq:t0t3}) for the shown representative values of $(t_0,t_3)$ (in units of MeV~fm$^3$ and MeV~fm$^6$, respectively) compared with data (``Exp.") where available and with results obtained with the Gogny D1S interaction (``D1S").  Our results at these chosen values of the parameters appear with our other results in Fig.~\ref{Fig:EcaD} along with the results from experiment and from the Gogny D1S interaction.
For references to data see Table~\ref{Tab:Data}. \label{Tab:EcaD}}
\end{table} 

\section{Summary and prospects\label{Sec:Concl}} 

The Daejeon16 two-nucleon interaction was employed in many-body approaches based on the mean-field approximation. 
The perturbative nature of Daejeon16 was verified by comparing 1) HF results with NCSM results for $^{16}$O and 2) the magnitude of perturbative corrections in light and heavy nuclei. 
Generally the heavier nuclei are obtained as overly dense and compressed. 
A phenomenological correction in the form of a two-plus-three-nucleon contact interaction was introduced in order to describe energies and radii across the nuclear chart within HF. 
The simultaneous good description of isospin-symmetric nuclei $^{16}$O and $^{40}$Ca along a region of only two weak and interdependent phenomenological terms confirms the already 
good optimization of the Daejeon16 interaction.
With the selected parameters we achieved a good description of the dipole poarizability in both $^{48}$Ca and $^{208}$Pb. 
The present results provide further justification for the use of Daejeon16 augmented with phenomenological corrections as an effective interaction of perturbative character in a variety of applications.

A simultaneous good description of both medium-mass and heavy nuclei, as well as light nuclei, 
could not be achieved with the present correction term. 
The parameters of the phenomenological correction require some fine tuning depending on the mass region. 
In order to improve on the present results one could consider an even richer density dependence for the phenomenological correction, for example, 
including higher powers of the density. 
Even so, the relative weakness of the correction needed to achieve reasonable saturation properties
with the Daejeon16 NN interaction suggests the finely-tuned nature of nuclear saturation.


\section*{Acknowledgments} 
Discussions with Fritz Coester during the early stages of this work are gratefully acknowledged. 
We wish to thank Ik Jae Shin for help with the Daejeon16 matrix elements. 
The work of P.P. and Y.K. was supported  by the Rare Isotope Science Project of the Institute for Basic Science funded 
by Ministry of Science, ICT and Future Planning and the National Research Foundation (NRF) of Korea (2013M7A1A1075764).
The work of J.P.V. was supported in part by the U.S. Department of Energy under Grants No. DE-FG02-87ER40371 and No. DE-SC0018223 (SciDAC-4/NUCLEI).

\end{document}